\documentclass[journal]{IEEEtran}
\usepackage{graphicx}
\ifCLASSINFOpdf
\else
\fi
\usepackage{amsmath}
\begin{document}
%
\title{Quaternion-Valued Single-Phase Model for Three-Phase Power Systems}
%
%
%

\author{Xiaoming~Gou,
        Zhiwen~Liu,
        Wei~Liu,~\IEEEmembership{Senior~Member,~IEEE,}
        Yougen~Xu,
        and~Jiabin~Wang,~\IEEEmembership{Senior~Member,~IEEE}
\thanks{This work was supported in part by the National Natural Science Foundation of China (61331019, 61490691).}
\thanks{X. Gou, Z. Liu, Y. Xu are with the School of Information and Electronics, Beijing Institute of Technology, Beijing 100081, China (email: xmgou@bit.edu.cn, zwliu@bit.edu.cn, yougenxu@bit.edu.cn).}
\thanks{W. Liu and J. Wang are with the Department of Electronic and Electrical Engineering, University of Sheffield, Sheffield S1 3JD, United Kingdom (email: w.liu@sheffield.ac.uk, j.b.wang@sheffield.ac.uk).}
}

%
%

\markboth{Journal of \LaTeX\ Class Files,~Vol.~14, No.~8, August~2015}%
{Shell \MakeLowercase{\textit{et al.}}: Bare Demo of IEEEtran.cls for IEEE Journals}
%



\maketitle

\begin{abstract}
 In this work, a quaternion-valued model is proposed in lieu of the Clarke's $\alpha$, $\beta$ transformation to convert three-phase quantities to a hypercomplex single-phase signal. The concatenated signal can be used for harmonic distortion detection in three-phase power systems. In particular, the proposed model maps all the harmonic frequencies into frequencies in the quaternion domain, while the Clarke's transformation-based methods will fail to detect the zero sequence voltages. Based on the quaternion-valued model, the Fourier transform, the minimum variance distortionless response (MVDR) algorithm and the multiple signal classification (MUSIC) algorithm are presented as examples to detect harmonic distortion. Simulations are provided to demonstrate the potentials of this new modeling method.
\end{abstract}


%
\IEEEpeerreviewmaketitle

\section{Introduction}
Power quality control is one of the major concerns for power delivery systems to function reliably, and it requires measurements of voltage characteristics, among which the frequency measurement is a non-trivial task due to the presence of voltage sags and voltage harmonics mostly caused by nonlinear loads \cite{BollenM2000}. In the particular case of three-phase power systems, the Clarke's $\alpha$, $\beta$ transformation is widely used as the preprocessing method to create a complex-valued single-phase signal from the real-valued three-phase signals \cite{AkkeM1997}, so that traditional complex-valued spectrum estimation methods can be applied, such as the MVDR method or the recently proposed Iterative MVDR (I-MVDR) method \cite{JeonH2010,XiaY2013}. To improve the resolution, we can further apply the subspace methods and one representative example is the MUSIC method \cite{SchmidtR1986}.

However, all the zero sequence voltages will be cancelled out in the complex-valued signal and hence can not be detected. Although these harmonic voltages would simply be blocked by a delta transformer, they will add up in the neutral, leading to overheating in the transformer and potential fire hazards \cite{GradyW2001}. To detect these harmonics, as well as harmonics of other orders, we propose a quaternion-valued model and all the traditional spectrum estimation methods can be extended to this domain, such as MVDR and MUSIC. We will show that harmonics of all orders will be reserved in the resulting quaternion-valued signal and will be detected by relevant estimation methods.

The rest of this paper is organised as follows. A brief review of the complex-valued model is presented in Section II. Our quaternion-valued model is proposed in Section III, together with the Fourier analysis and the MVDR and MUSIC-like estimation algorithms. Simulation results are provided in Section IV and conclusions are drawn in Section V.

\section{Complex-valued frequency estimation for three-phase power systems}
\subsection{A brief review}
We consider the following discrete-time balanced three-phase power system in the presence of $H-1$ harmonic distortions:
\begin{equation}
\begin{split}
v_a(n)&=\sum_{h=1}^HV_h\cos(h(\Omega nT_s+\phi))+\epsilon_a(n)\;,\\
v_b(n)&=\sum_{h=1}^HV_h\cos\Big(h\Big(\Omega nT_s+\phi-\frac{2\pi}{3}\Big)\Big)+\epsilon_b(n)\;,\\
v_c(n)&=\sum_{h=1}^HV_h\cos\Big(h\Big(\Omega nT_s+\phi+\frac{2\pi}{3}\Big)\Big)+\epsilon_c(n)\;,
\end{split}
\end{equation}
where $\{V_h\}_{h=1}^H$ are the amplitudes of the harmonic signals, $\Omega$ is the fundamental (angular) frequency, $T_s$ is the sampling interval, $\phi$ is the signal phase, and $\epsilon_a(n),\epsilon_b(n),\epsilon_c(n)$ are the measurement noise.

Traditionally, the three-phase signals will be converted to a complex-valued single-phase signal via the Clarke's $\alpha,\beta$ transformation. Firstly, the three-phase signals are mixed into two parts, namely $v_\alpha(n)$ and $v_\beta(n)$, where
\begin{equation}
\begin{bmatrix}
v_\alpha(n)\\
v_\beta(n)
\end{bmatrix}
=\mathbf T
\begin{bmatrix}
v_a(n)\\
v_b(n)\\
v_c(n)
\end{bmatrix}
\;,
\end{equation}
\begin{equation}
\mathbf T=\frac{2}{3}
\begin{bmatrix}
1 & -\frac{1}{2} & -\frac{1}{2}\\
0 & \frac{\sqrt 3}{2} & -\frac{\sqrt 3}{2}
\end{bmatrix}
\;.
\end{equation}
Then these two parts will be merged as a complex-valued signal $v_{cv}(n)=v_\alpha(n)+\imath v_\beta(n)$.

With this complex-valued signal, we can exploit the MVDR spectrum to locate the frequencies, and it is given by:
\begin{equation}
S_{\mathrm {MVDR}}(\Omega)=\frac{1}{\mathbf s^\mathrm H(\Omega)\mathbf R^{-1}\mathbf s(\Omega)}\;,\label{MVDR}
\end{equation}
where $(\cdot)^\mathrm H$ is the Hermitian-transpose operation, $\mathbf R$ is the covariance matrix of dimension $M\times M$, and
\begin{equation}
\mathbf s(\Omega)=[1,e^{-\imath\Omega T_s},\cdots,e^{-\imath\Omega T_s(M-1)}]^\mathrm T
\end{equation}
is the frequency sweeping vector.

We can also use the MUSIC spectrum which is expressed as:
\begin{equation}
S_{\textrm{MUSIC}}(\Omega)=\frac{1}{||\mathbf s^\mathrm H(\Omega)\mathbf U_N||^2}\;,\label{MUSIC}
\end{equation}
where $||\cdot||$ denotes the Euclidean norm, $\mathbf U_N$ represents the noise subspace and comprises the eigenvectors of the covariance matrix $\mathbf R$ which are corresponding to the $M_0$ smallest eigenvalues, where $M_0$ is assumed to be known or can be estimated using the information theory methods \cite{WaxM1985}.

In practice, the covariance $\mathbf R$ needs to be updated and estimated from the average of samples,
\begin{equation}
\hat{\mathbf R}(n)=\mathbf V_{cv}(n)\mathbf V_{cv}^\mathrm H(n)/K\;,
\end{equation}
where
\begin{equation}
\mathbf V_{cv}(n)=\!
\begin{bmatrix}
v_{cv}(n) & \cdots & v_{cv}(n-K+1)\\
\vdots & \ddots & \vdots\\
v_{cv}(n-M+1) & \cdots & v_{cv}(n-K-M+2)
\end{bmatrix}
\;,
\end{equation}
where $K$ is the number of observations.

\subsection{Missing harmonic signals in the complex-valued signal}
In detail, $v_{cv}(n)$ is composed of complex-domain harmonic signals that can be divided into two categories plus noise,
\begin{equation}
v_{cv}(n)=v_{cv1}(n)+v_{cv2}(n)+\epsilon_{cv}(n)\;,
\end{equation}
where $v_{cv1}(n)$ is the summation of all positive sequence voltages,
\begin{equation}
v_{cv1}(n)=\sum_{p=1}^{\lfloor\frac{H+2}{3}\rfloor}V_{3p-2}e^{\imath(3p-2)(\Omega nT_s+\phi)}\;,
\end{equation}
and $v_{cv2}(n)$ is the summation of all negative sequence voltages,
\begin{equation}
v_{cv2}(n)=\sum_{p=1}^{\lfloor\frac{H+1}{3}\rfloor}V_{3p-1}e^{-\imath(3p-1)(\Omega nT_s+\phi)}\;,
\end{equation}
and $\lfloor x\rfloor$ denotes the largest integer not greater than $x$. All the zero sequence voltages have been cancelled out. Zero sequence voltages of the same order are cophasial in the three voltage channels and will be eliminated since both rows of the transformation matrix $\mathbf T$ are zero-mean vectors.

To solve this problem, we propose our quaternion-valued approach in the next section.

\section{Quaternion-valued frequency estimation for three-phase power systems}

We construct a quaternion-valued signal from the three-phase signals as
\begin{equation}
v_{qv}(n)=\imath v_a(n)+\jmath v_b(n)+kv_c(n)\;,
\end{equation}
where $\imath,\jmath,k$ are the three imaginary units of the quaternion algebra which are constrained by \cite{WardJ1997}
\begin{equation}
\begin{split}
\imath^2&=\jmath^2=k^2=\imath\jmath k=-1\;,\\
\imath\jmath&=-\jmath\imath=k\;,\\
\jmath k&=-k\jmath=\imath\;,\\
k\imath&=-\imath k=\jmath\;.
\end{split}
\end{equation}
This quaternion-valued signal contains quaternion-domain harmonic signals that belong to three categories $v_{qv1}(n)$, $v_{qv2}(n)$, $v_{qv3}(n)$ plus noise,
\begin{equation}
\begin{split}
v_q(n)&=\imath v_a(n)+\jmath v_b(n)+kv_c(n)\\
&=v_{qv1}(n)+v_{qv2}(n)+v_{qv3}(n)+\epsilon_{qv}(n)\;,
\end{split}
\end{equation}
where $v_{qv1}(n)$ is the summation of all the positive sequence voltages,
\begin{equation}
\begin{split}
v_{qv1}(n)&=\sum_{p=1}^{\lfloor\frac{H+2}{3}\rfloor}\frac{2\imath-\jmath-k}{2}V_{3p-2}\big\{\cos[(3p-2)(\Omega nT_s+\phi)]\\
&\quad-\frac{\imath+\jmath+k}{\sqrt{3}}\sin[(3p-2)(\Omega nT_s+\phi)]\big\}\\
&=\sum_{p=1}^{\lfloor\frac{H+2}{3}\rfloor}\frac{2\imath-\jmath-k}{2}V_{3p-2}e^{-\frac{\imath+\jmath+k}{\sqrt{3}}(3p-2)(\Omega nT_s+\phi)}\;,
\end{split}
\label{3p-2}
\end{equation}
$v_{qv2}(n)$ is the summation of all the negative sequence voltages,
\begin{equation}
\begin{split}
v_{qv2}(n)&=\sum_{p=1}^{\lfloor\frac{H+1}{3}\rfloor}\frac{2\imath-\jmath-k}{2}V_{3p-1}\big\{\cos[(3p-1)(\Omega nT_s+\phi)]\\
&\quad+\frac{\imath+\jmath+k}{\sqrt{3}}\sin[(3p-1)(\Omega nT_s+\phi)]\big\}\\
&=\sum_{p=1}^{\lfloor\frac{H+1}{3}\rfloor}\frac{2\imath-\jmath-k}{2}V_{3p-1}e^{\frac{\imath+\jmath+k}{\sqrt{3}}(3p-1)(\Omega nT_s+\phi)}\;,
\end{split}
\label{3p-1}
\end{equation}
$v_{qv3}(n)$ is the summation of all the zero sequence voltages,
\begin{equation}
\begin{split}
v_{qv3}(n)&=\sum_{p=1}^{\lfloor\frac{H}{3}\rfloor}V_{3p}(\imath+\jmath+k)\cos[3p(\Omega nT_s+\phi)]\\
&=\sum_{p=1}^{\lfloor\frac{H}{3}\rfloor}V_{3p}\frac{\imath+\jmath+k}{2}\Big(e^{\frac{\imath+\jmath+k}{\sqrt{3}}3p(\Omega nT_s+\phi)}\\
&+e^{-\frac{\imath+\jmath+k}{\sqrt{3}}3p(\Omega nT_s+\phi)}\Big)\;.
\end{split}
\label{3p}
\end{equation}
Hence all the harmonic signals will be reserved in the quaternion-valued signal. We may observe from (\ref{3p-2})--(\ref{3p}) that the frequencies of the harmonic signals have been mapped into the frequencies of the quaternion-valued signal associated with the $\frac{\imath+\jmath+k}{\sqrt 3}$ axis (see Fig. 1).
\begin{figure}
  \centering
  \includegraphics[width=2.2in]{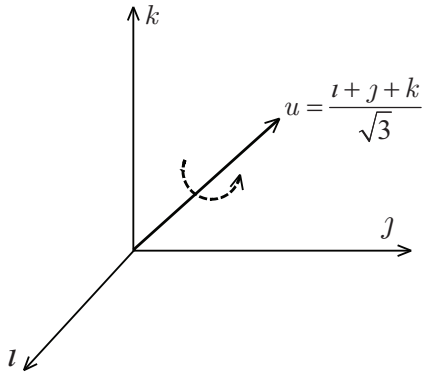}\\
  \vspace{.5cm}
  \caption{Quaternion domain frequency associated with the $\frac{\imath+\jmath+k}{\sqrt 3}$ axis}
\end{figure}
\begin{figure}[h]
  \vspace{.5cm}
  \centering
  \includegraphics[width=3.2in]{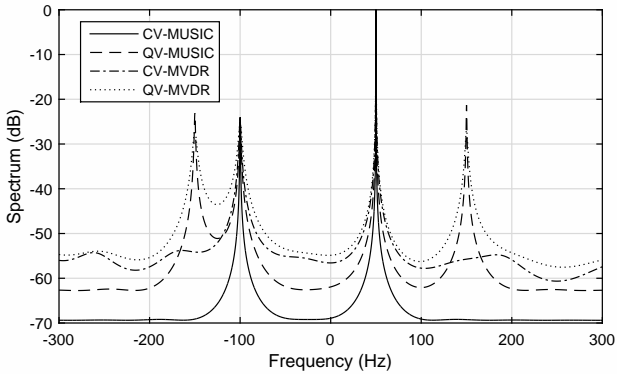}\\
  \caption{MVDR and MUSIC spectra of quaternion- and complex-valued models}
\end{figure}

Then we can adopt the MVDR spectrum in (\ref{MVDR}) and the MUSIC spectrum \footnote{Not to be confused with the Quaternion-MVDR (Q-MVDR) algorithm \cite{TaoJW2013} for the adaptive beamforming with vector-sensor array beamforming or the Quaternion-MUSIC (Q-MUSIC) algorithm \cite{MironS2006} for the direction-of-arrival estimation with vector-sensor arrays. Their steering vectors are complex-valued vectors multiplied by quaternion-valued scalars, which are conceptionally different from the quaternion-valued frequency sweeping vector defined in this paper. We marked our algorithms by QV-MVDR and QV-MUSIC for clarification.} in (\ref{MUSIC}) by substituting the frequency sweeping vector as
\begin{equation}
\mathbf s(\Omega)=[1,e^{-\frac{1}{\sqrt{3}}(\imath+\jmath+k)\Omega T_s},\cdots,e^{-\frac{M-1}{\sqrt{3}}(\imath+\jmath+k)\Omega T_s}]^\mathrm T\;.
\end{equation}

The frequencies detected in the spectrum are either the original real-domain angular frequencies or their additive inverses, namely
\begin{enumerate}
\item[(1)] If a peak is detected in the spectrum in the absence of its additive inverse, it corresponds to a positive or negative sequence voltage signal and this spectrum peak indicates its angular frequency or its additive inverse.
\item[(2)] If two ``mirrored'' peaks are detected in the spectrum, they correspond to a zero sequence voltage signal and they indicate the signal's angular frequency and its additive inverse, respectively.
\end{enumerate}

\begin{figure}
\vspace{-2cm}
\centering
\begin{minipage}[b]{\linewidth}
  \centering
  \centerline{\includegraphics[width=3.8in]{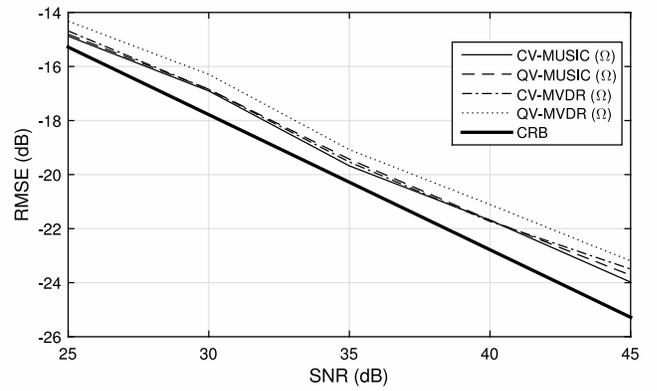}}
  \vspace{-6.5cm}
  \centerline{\scriptsize{(a) Fundamental frequency}}
\end{minipage}
\hfill
\begin{minipage}[b]{\linewidth}
  \centering
  \centerline{\includegraphics[width=3.8in]{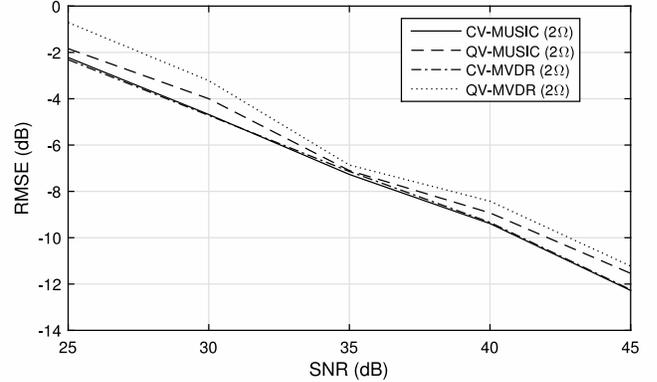}}
  \vspace{-6.5cm}
  \centerline{\scriptsize{(b) Second-order harmonic frequency}}
\end{minipage}
\begin{minipage}[b]{\linewidth}
  \centering
  \centerline{\includegraphics[width=3.8in]{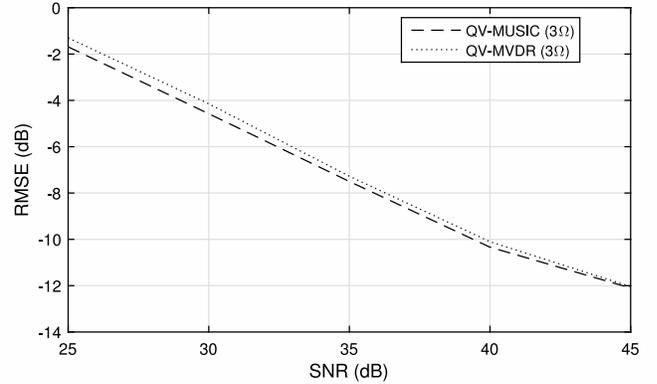}}
  \vspace{-6.5cm}
  \centerline{\scriptsize{(c) Third-order harmonic frequency}}
\end{minipage}
\caption{Averaged estimation errors of quaternion- and complex-valued algorithms}
\end{figure}

\section{Simulations}

In this section, we provide some numerical examples to illustrate the performance of the proposed quaternion model. In all experiments, the fundamental frequency is 50 Hz, the sampling frequency is $f_s=1/T_s=20$ KHz, the initial phase is $\phi=\pi/7$, and $K=80$, $M=32$. There exist a second-order and a third-order harmonic signals, both set to be 6\% in amplitude.

In the first experiment, we test the capability of the two modelings. The MVDR and MUSIC spectra of the quaternion- and complex-valued models are plotted in Fig. 2, where SNR $=40$ dB. It can be observed that the proposed model is able to detect all the harmonic signals, namely 50 Hz (the fundamental frequency), $-100$ Hz (the second-order harmonic), and $\pm$150 Hz (the third-order harmonic), while the complex-valued model fails at the third-order harmonic frequency.

In the second experiment, we test the accuracy of relevant algorithms. The estimation errors (averaged via 300 Monte Carlo simulation runs) of the quaternion- and complex-valued MVDR and MUSIC algorithms are plotted in Fig. 3, where the SNR value varies from 25 to 45 dB. We can see that all the algorithms have similar estimation accuracy.


\section{Conclusion}
\label{sec:prior}

We have presented a quaternion-valued model as an alternative preprocessing approach to convert the three-phase signals into a single-phase system. Compared with the Clarke's transformation, the proposed model can additionally detect the zero sequence voltages. Simulated results show that the proposed model can detect all-order voltage harmonics effectively, while sharing similar estimation accuracy with the complex-valued model.



\ifCLASSOPTIONcaptionsoff
  \newpage
\fi

\end{document}